\documentclass[preprint,12pt,authoryear]{elsarticle}
\usepackage{subcaption}
\usepackage{xcolor}
\usepackage{soul}
\usepackage[most]{tcolorbox}
\usepackage{amssymb}

\usepackage{amsmath}

\journal{Atmospheric Research}

\begin{document}

\begin{frontmatter}

\title{Automatic detection of overshooting tops and their properties from visible satellite channels} 

\author[1]{Anežka Doležalová}
\author[2]{Jakub Seidl}
\author[2]{Jindřich Šťástka}
\author[3]{Ján Kaňák}

\affiliation[1]{organization={Department of Atmospheric Physics, Faculty of Mathematics and Physics, Charles University},
            addressline={V Holešovičkách 747/2}, 
            city={Prague},
            postcode={180 00}, 
            country={Czech Republic}}
\affiliation[2]{organization={Czech Hydrometeorological Institute}, 
            city={Prague},
            country={Czech Republic}}
\affiliation[3]{organization={Slovak Hydrometeorological Institute},
            city={Bratislava},
            country={Slovakia}}

\begin{abstract}
Overshooting tops (OTs) are informative indicators of convective storm intensity and are widely utilized in meteorological analyses. This study presents an automated algorithm for OT detection and OT height estimation using convolutional neural networks applied to visible satellite imagery. The models are trained and validated on an extensive OT dataset comprising approximately 10,000 manually detected cases over Europe. The OTs were identified from high-resolution visible (HRV) channel of the SEVIRI instrument on board the MSG geostationary satellite, with the heights determined from the length of their shadows in the imagery. While conventional OT detection methods primarily rely on the identification of cold features in thermal infrared channels, our approach extracts information from visible channels, leveraging the ground truth data on OT shadow length provided by the training dataset. In the morning and afternoon hours, when the shadows are visible, the proposed models detect OTs with a probability of detection reaching 97\% and estimate their height with an average error of 0.25 km. The performance is expected to further improve once the model is applied to polar and new generation geostationary satellites with increased spatial resolution.

\end{abstract}

\begin{graphicalabstract}
\includegraphics[width=1\linewidth]{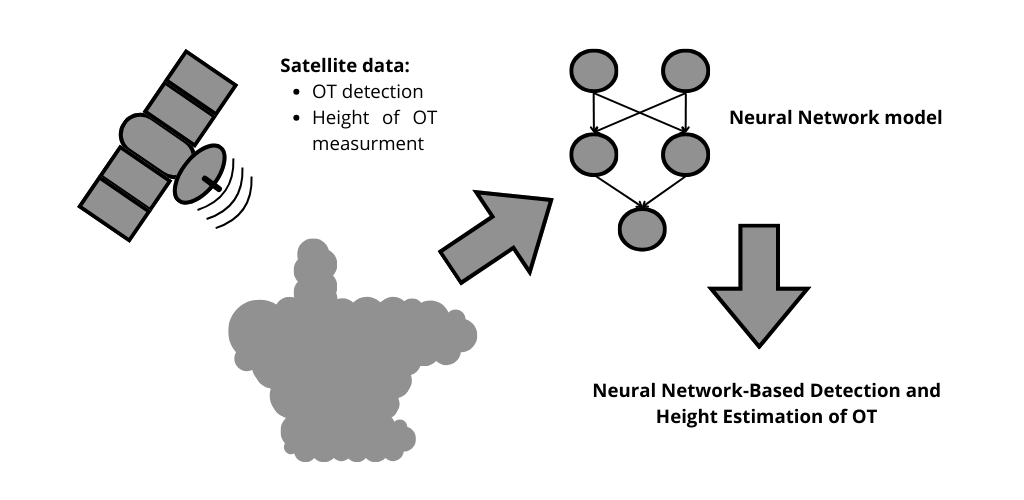}
\end{graphicalabstract}

\begin{highlights}
\item High-Accuracy OT detection using Visible Channel and Neural Network
\item Operational product development and Real-World applicability
\item OT height accurately estimated using visible imagery and shadow geometry via Neural Network
\end{highlights}


\end{frontmatter}

\section{Introduction}

Convection and associated convective storms are a significant source of extreme weather events, often leading to hazardous conditions such as hail, tornadoes, strong winds, and flooding. These phenomena can cause extensive damage to both property and human life worldwide. At the cloud tops of such storms, various distinct features can be observed in satellite imagery. These features are strongly correlated with storm severity \citep{setv2010, RICCHI}, making their study essential for improving forecasting and risk assessment. In this work, we focus on overshooting tops (OTs), which form when a strong updraft penetrates the tropopause. Previous research on OT height determination has been conducted, among others, by Kristopher Bedka, who developed several methodologies of OT detection \citep{bedka2010}, and Ján Kaňák, whose method and database serve as a foundation for our study \citep{kanak2012}. The direct relationship between OT height and updraft strength and position has been found, for example, by \citet{Bluestein2019}. Other studies that found a relationship between properties of OT and severe weather include \citet{dworak2012} and \citet{sun2024}.These studies used satellite data, which is the most suitable for observing this feature, but it can also be detected using other methods, such as radar or other ground-based measurements \citep{YAIR}.

Most existing OT detection approaches rely on thermal infrared (IR) satellite channels, either by analyzing single-channel temperature variations or using differences between multiple IR channels. Various methodologies have been tested in \citet{mikus}, with additional validation conducted on the database presented in \citet{rad2015}. The most effective of these approaches, the O3-IRW method, detected only 61\% of OTs in the \citet{rad2015} dataset. The O3-IRW method detects overshooting tops by identifying positive brightness temperature differences between the ozone absorption channel (9.7 µm) and the infrared window channel (10.8 µm), which may highlight stratosphere-penetrating cloud tops due to their distinct thermal and compositional signatures. The primary limitation of these techniques lies in their reliance on temperature-based thresholds, as OTs may exhibit temperature anomalies that complicate detection. 

The most widely used and cited technique for overshooting top (OT) detection was developed by Kristopher Bedka. Known as the Infrared Window (IRW)-Texture Technique, it utilizes brightness temperature gradients within the infrared window channel to identify OTs. The method combines several key components: (1) brightness temperature gradients, where colder pixels associated with OTs are detected relative to the surrounding anvil cloud; (2) numerical weather prediction (NWP) tropopause temperature forecasts, which help distinguish cloud tops that penetrate into the stratosphere; and (3) criteria for OT size and brightness temperature, derived from the analysis of 450 thunderstorm events. These criteria define the spatial and thermal characteristics typical of overshooting tops. The false alarm rate of the IRW-texture method ranges from 4.2\% to 38.8\%, depending on the strength of the overshooting event and the specific quality control settings applied in the algorithm \citep{bedka2010}.

Beyond traditional threshold-based approaches, deep learning techniques have also been applied to OT detection, as demonstrated in \citet{kan2020} and \citet{kim2018}. These studies utilized both visible and thermal satellite channels, achieving Probability of Detection (PoD) values of 79.31\% for GOES satellite data and 79.68\% for tropical regions using Himawari satellite data. The effectiveness of deep learning models is highly dependent on dataset size, with the GOES study utilizing only 459 images, while the Himawari dataset contained 10,000 training samples and a total of 571 OT cases.  

Despite numerous studies on OT detection, relatively little research has been conducted on OT height estimation, which is primarily feasible using visible satellite imagery. One such study is \citet{kanak2012}, whose methodology we adopt in this work. Another relevant study is that of \citet{Bluestein2019}, which demonstrated that colder OT temperatures (indicating higher cloud tops) and larger OT areas are strongly associated with hail and severe winds, thereby improving storm intensity nowcasting. The model developed in our work, which performs both joint OT detection and height estimation, could enable further statistical investigation into the relationship between OT height, storm dynamics, and the occurrence of hazardous phenomena.

The presence of OTs serves as one of the important indicators for meteorologists \citep{pia}, signaling the potential severity of a convective storm. Our objective is to develop an automated model for OT detection using satellite data, providing a tool that highlights OTs for operational use by forecasters and other applications. Moreover, beyond mere OT presence, additional properties such as height, width, and lifetime can be derived from satellite imagery—particularly from visible channels—and analyzed for their significance.

Our approach to comprehensive OT detection and property evaluation is based on machine learning techniques, specifically neural networks. The models were trained on an extensive dataset containing approximately 10,000 OT cases over Europe. Given that cloud-top features, including OTs, are most effectively observed via satellite imagery, we utilized data from the high-resolution visible (HRV) channel of the SEVIRI instrument onboard MSG satellites. Our models can accurately detect the position of OTs within satellite imagery and estimate their height using the same data source.  

The structure of this paper is as follows. In Section 2, we introduce the theoretical background, describe the satellite data used, and explain the machine learning models developed for overshooting top detection and height estimation. Section 3 presents the results: first, the performance of the OT detection models, then the regression model for estimating OT heights, and finally a comparison with other existing methods. The paper concludes in Section 4, where we summarize the key findings and suggest directions for future research.

\section{Theory, Data, and Methods}

Overshooting tops (OTs) are cloud-top phenomena associated with convective storms and they can be characterized by multiple properties. In this study, we primarily focus on the detection of OTs and the estimation of their height. The methodology for measuring OT height follows the approach developed by \citet{kanak2012}, which utilizes satellite observations in the visible spectrum. In these images, OTs cast shadows onto the surrounding cloud top, which can be measured to determine their height above the cloud top.  

\begin{figure}[ht!]
\centering
\includegraphics[width=0.99\linewidth]{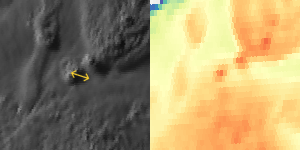}
\caption{Example of OT measurement using visible and infrared satellite channels. \textbf{Left:} High-resolution visible (HRV) image showing the OT with its shadow (yellow arrow shows one of the examples), which is used for height estimation. \textbf{Right:} Corresponding infrared (IR) brightness temperature image (specifically product "colorized ir clouds"), highlighting the colder OT region.}
\label{fig:example_OT_measurement}
\end{figure}

An example of OT and its height visualization in both visible and infrared satellite channels is presented in Figure~\ref{fig:example_OT_measurement}. The left panel shows a high-resolution visible (HRV) image, where the OT casts a distinct shadow (indicated by the yellow arrow), allowing for height estimation based on shadow length and sun elevation angle. The right panel displays the corresponding infrared (IR) brightness temperature image (specifically product "colorized ir clouds") commonly used in OT detection due to the colder temperatures of overshooting tops compared to their surroundings.

\begin{figure}[ht]
\centering
\includegraphics[width=0.7\linewidth]{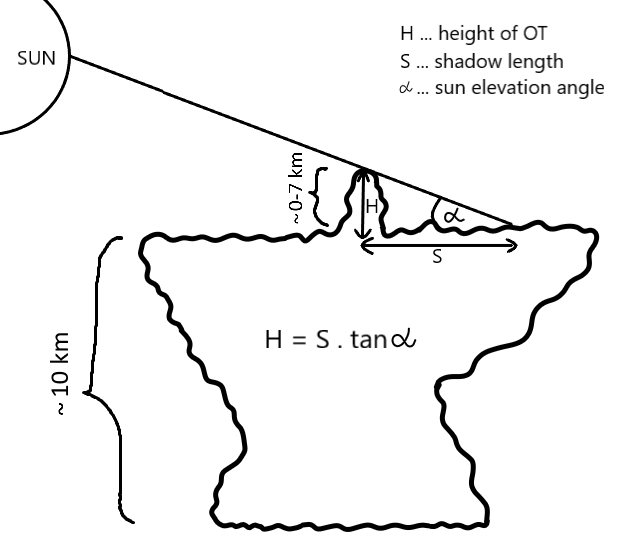}
\caption{Schematic representation of the geometric approach used for OT height estimation. The shadow length $S$ is measured in HRV satellite images, and the height $H$ is computed using the sun elevation angle $\alpha$.}
\label{fig:geometry}
\end{figure}

The relationship between the observed shadow length $S$ and the OT height $H$ follows a simple geometric principle, as illustrated in Figure~\ref{fig:geometry}:  

\begin{equation}
\label{eq:shadow_length_to_height}
    H = S \cdot \tan \alpha,
\end{equation}  

where $S$ is the shadow length measured in high-resolution visible (HRV) satellite images, and $\alpha$ is the sun elevation angle.  

When using the sun elevation angle for OT height estimation, a minor difference exists between the sun’s angle at the Earth’s surface and at cloud-top level. However, this difference is minimal—typically around 1\textperthousand\ when comparing projections from the Earth's surface to an altitude of 10 km, which is the common height of the tropopause in the region of Europe where our study is located. This negligible discrepancy allows for a simplified approximation without introducing significant errors. In practice, we assume an average cloud-top height of 10 km, acknowledging that actual values vary from storm to storm but remain within a reasonable range for this estimation method.  

The sun elevation angle $\alpha$ plays a crucial role in measuring OT height from shadow length. It can be incorporated in two ways.
Postprocessing approach, which means that the model predicts only the shadow length, and the OT height is computed using the sun elevation angle afterward.  
Direct input approach means that the sun elevation angle is included as an input feature in the model, allowing it to predict OT height directly from satellite images.

For this study, we adopt the first approach, in which the sun elevation angle is utilized in postprocessing. This simplifies the model architecture and improves interpretability while maintaining accurate height estimates.  

\begin{figure}[ht!]
\centering
\includegraphics[width=0.7\linewidth]{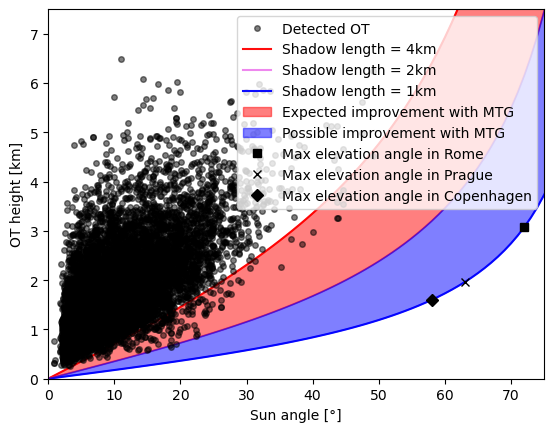}
\caption{Relationship between OT height and sun elevation angle in the database. Black dots represent detected OTs, with shadow-based height estimates. The red, purple, and blue lines correspond to theoretical OT height estimates based on different shadow lengths (4 km, 2 km, and 1 km, respectively). The shaded regions indicate potential improvements in OT detection due to improved spatial resolution achievable with the upcoming MTG satellite mission. The three unique symbols on the right mark the maximum sun elevation angles at various locations (Rome, Prague, and Copenhagen).}
\label{fig:sunangle}
\end{figure}

OT shadows are detectable in visible satellite channels that rely on sunlight reflection from the Earth and clouds. However, the diurnal cycle imposes certain limitations. Around midday, the sun may be positioned nearly overhead, causing shadows to be minimal or even undetectable, which reduces the effectiveness of OT height estimation and even detection during this part of the day. Furthermore, since visible channels depend on sunlight, they cannot be used for OT detection at night. Figure~\ref{fig:sunangle} demonstrates the relationship between the sun elevation angle and the height of the OTs detected by human experts in the HRV channel of the MSG satellite, emphasizing the limitations imposed by the diurnal cycle and the satellite pixel resolution. The figure reveals empirical thresholds for reliable OT shadow detection. When the shadow length falls below approximately 4 km, human observers find reliable detection increasingly challenging, resulting in a significant decline in the frequency of detected OTs with shorter shadows. A secondary threshold is observed at roughly 2 km, below which no OT shadows have been detected. These resolution-based thresholds translate into minimum OT heights required for reliable or feasible detection at specific sun angles. At any given latitude, these thresholds further define the portion of the day during which OT detection is reliable or possible. Compared to MSG, polar-orbiting and next-generation geostationary satellites offer enhanced spatial resolution, thereby extending this detection window. For MTG, the expected improvement is indicated in Figure~\ref{fig:sunangle} by the blue and red shaded areas, illustrating the reduced minimum OT height detectable at specific sun zenith angles. For example, at a sun zenith angle of 30°, the minimum OT height for reliable shadow detection is expected to decrease from approximately 2 km (MSG) to about 1 km (MTG), with potential detection capability extending down to 0.5 km. Further inaccuracy can be caused by situations where the shadow is long and extends beyond the cloud top area and due to observer inaccuracy, which is approximately less than 10\%.

\subsection{Database}

\begin{figure}[ht!]
\centering
\includegraphics[width=0.7\linewidth]{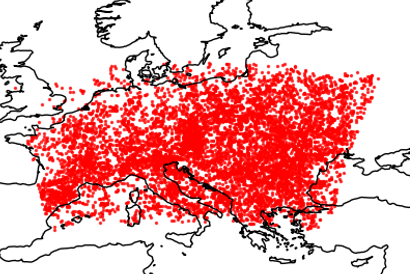}
\caption{Geographical distribution of detected overshooting tops (OTs) within the database. The dataset consists of approximately 10,000 OT cases identified over Europe using SEVIRI instrument data from MSG satellites.}
\label{fig:map}
\end{figure}

In \citet{kanak2012}, Ján Kaňák assembled a comprehensive database of overshooting tops (OTs) along with measurements of their shadow lengths. This dataset comprises approximately 10,000 OT cases over Europe (see coverage in Figure~\ref{fig:map}) from years 2009-2014. The database was created using observations from the SEVIRI instrument onboard the MSG satellite series. Each OT was manually identified and measured using specialized software, employing the Albers projection with a spatial resolution of 1 km per pixel \citep{kanak2012}. The database contains coordinates of the OT centers together with the length of its shadow (in kilometers) and the OT height computed using Eq. (\ref{eq:shadow_length_to_height}).

For model training in Section \ref{sec:results}, each OT from the database is represented by an HRV image in the Azimuthal Equidistant projection, which ensures uniform pixel size - 1 km per pixel in our case, centered at the OT coordinates. This projection compensates for the varying native resolution of satellite data across the field of view, particularly the resolution decrease toward the north \citep{schmetz2002}. As a result, the model consistently interprets each input pixel as representing 1 × 1 km, regardless of the geographic location of the OT. Although the decreasing native resolution of satellite imagery toward the poles does not directly affect the model due to the projection used, in northern regions, neighboring pixels may contain identical or nearly identical values because of the coarser native resolution. This effect remains important for interpreting OT position and shadow length, and we recognize it as a genuine limitation of the data. The spatial resolution of the HRV channel gradually decreases in our region of interest from approximately 1.3 × 1.0 km in the southwest to about 1.9 × 1.06 km in the northeast.

In theory, the smallest shadow that can be detected in MSG satellite imagery corresponds to 1 km in nadir view (0.5 km for the upcoming MTG satellite series) and corresponding sizes in other positions. However, the significant limitations affect practical detection. Next to the desreasing resolution from nadir a shadow length of just 1 pixel ($\sim$1 km) is difficult to reliably distinguish from background brightness fluctuations, making shorter shadows prone to detection errors as was discussed in previous section and shown in Figure~\ref{fig:sunangle}.  

\begin{figure}[h]
\begin{subfigure}[t]{0.49\textwidth}
\centering
\includegraphics[width=\linewidth]{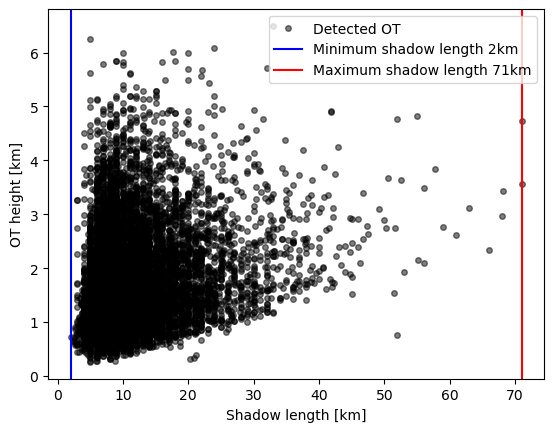}
\caption{Relationship between OT height and detected shadow length. Black dots represent OTs detected in the HRV channel of the SEVIRI instrument by human experts, blue and red line highlights the minimum and maximum value of detected shadow length, respectively.}
\label{fig:shadowlength}
\end{subfigure}
\hfill
\begin{subfigure}[t]{0.49\textwidth}
\centering
\includegraphics[width=\linewidth]{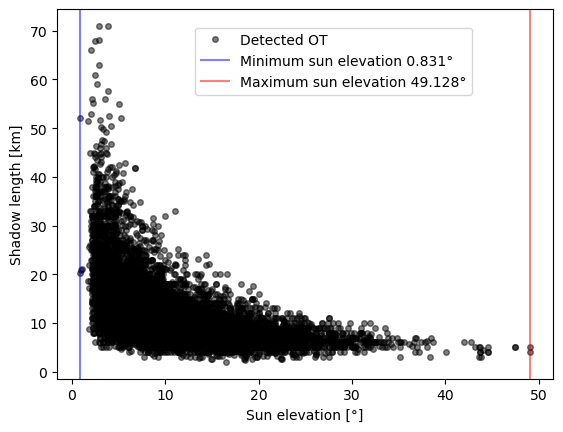}
\caption{Relationship between Sun elevation angle and detected OT shadow length. Black dots represent detected OTs, blue and red line highlights the minimum and maximum value of sun elevation angle.}
\label{fig:shadowangle}
\end{subfigure}
\caption{Comparison of OT shadow characteristics.}
\label{fig:both_shadows}
\end{figure}

As has already been said, the shortest shadow manually detected in the dataset was 2 km. Additionally, Figure~\ref{fig:shadowlength} demonstrates the relationship between OT height and detected shadow length, emphasizing the practical detection threshold. Shorter shadows (particularly those below 2 km) become increasingly difficult to distinguish from background variations, reducing their representation in the dataset. The trend also  shows that taller OTs generally cast longer shadows, making them more reliably detectable in HRV imagery. However, the frequency of the longest shadows also declines, partly because they occur at low sun angles in morning or evening, when low light often renders the scene too dark for reliable detection.

To support understanding, Figure~\ref{fig:shadowangle} illustrates the relationship between shadow length and sun elevation angle. The plot shows that longer shadows are associated with lower sun elevation angles, which is consistent with Eq. \ref{eq:shadow_length_to_height} - at lower sun angles, the same overshooting top casts a longer shadow compared to higher sun angles. The uncertainty of shadow length measurements increases with shadow length. This may result from the greater spatial extent itself or from low sun elevation conditions, during which the overall image brightness is reduced, making precise measurement more difficult. Nevertheless, the overall measurement uncertainty estimated by comparing measurements done by independent human experts remains below 10\% of the shadow length for all cases.

\subsection{Machine Learning Models} \label{ml}

Our task involves image recognition, specifically identifying distinct patterns in satellite data. Unlike conventional image recognition tasks that typically rely on RGB images (three channels), our approach selects one or more channels from a multispectral spectrometer. Given the complexity of the problem, we opted for neural networks, which are particularly well-suited for this type of pattern recognition. 

Artificial neural networks (NNs) are computational models inspired by the structure and function of the human brain. They are designed to process vast amounts of data and recognize intricate patterns and relationships. NNs consist of interconnected artificial neurons arranged in layers, with each connection assigned a specific weight. These layers are generally categorized as follows:
\begin{itemize}
\item The \textbf{input layer}, which receives the raw data.
\item The \textbf{hidden layers}, which process the data through a series of weighted connections and activation functions.
\item The \textbf{output layer}, which produces the final prediction.  
\end{itemize}

\begin{figure}[ht!]
\centering
\includegraphics[width=0.99\linewidth]{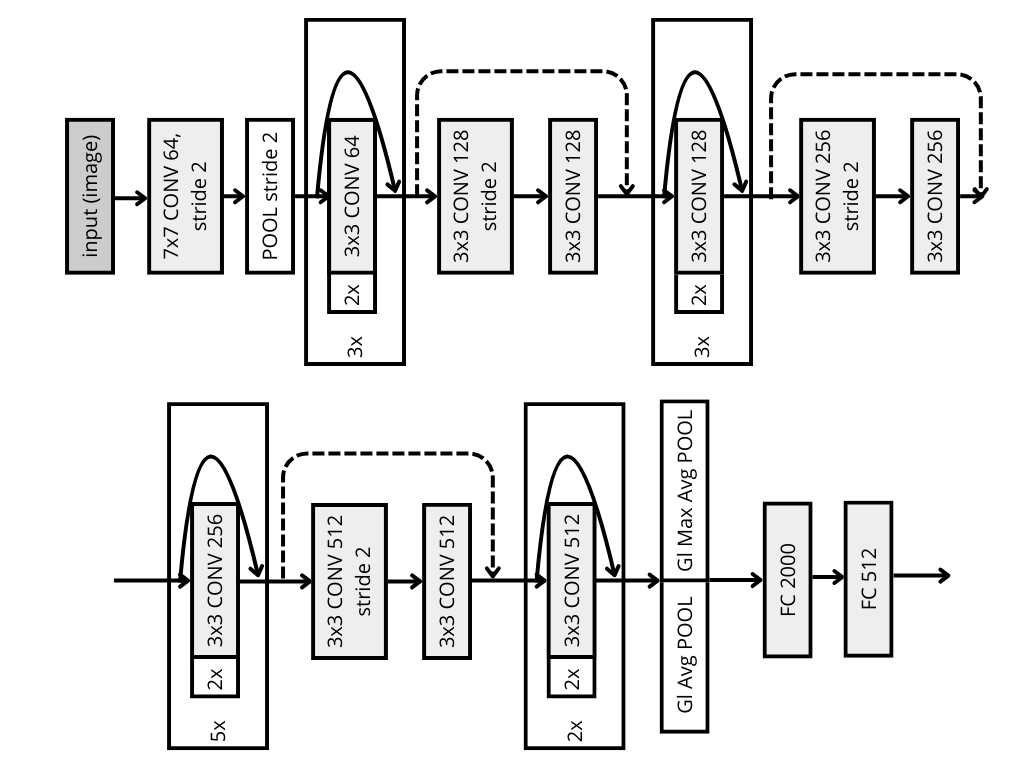}
\caption{Structure of the ResNet34 model, highlighting residual connections that improve training stability and accuracy \citep{resnet}. Abbreviations: convolution layer (CONV), fully connected layer (FC), global average pooling (Gl Avg POOL), global max average pooling (Gl Max Avg POOL).}
\label{fig:resnet}
\end{figure}

Various types of neural networks exist, but we focus on convolutional neural networks (CNNs), specifically the ResNet architecture \citep{resnet}. ResNet, or Residual Network, is a deep learning model incorporating residual connections to improve gradient flow and optimize training efficiency. The network begins with a convolutional layer, batch normalization, ReLU activation, and max pooling. It then passes through several stages of residual blocks, each containing two 3×3 convolutional layers with batch normalization. The skip connections add the input of each block directly to its output, allowing the model to learn residual functions. When dimensions differ, a 1×1 convolution is used in the shortcut path. The network ends with concatenated global average and max pooling followed by a regression/classification head consisting of two fully connected layers for OT classification or OT height regression. Apart from the initial max pooling after the first convolution, downsampling is performed by strided convolutions inside certain residual blocks as indicated in Figure~\ref{fig:resnet}. Our implementation utilizes ResNet18 and ResNet34, where the numbers refer to the depth of the network (i.e., the total number of layers 18 or 34). The specific structure of the ResNet34 architecture is illustrated in Figure~\ref{fig:resnet}. The model was trained using the Adam optimizer \citep{kingma2017}, and regularized with a dropout rate of 50\% used in the classification/regression head.

For implementation, we utilized the FastAI library \citep{Howard_Gugger_2021} in Python. The chosen ResNet models were pre-trained on the ImageNet \citep{imagenet} dataset, which comprises over 1.2 million images spanning 1,000 different classes (the categories most similar to the appearance of OT can be, for example, volcano, geyser, or alp). This pre-trained model enabled us to apply transfer learning \citep{weiss2016}, a technique where a model trained on a large dataset is fine-tuned on a smaller, domain-specific dataset—in this case, our database of overshooting tops (OTs). The original classification head was removed and replaced with two new fully connected layers tailored to the specific task—either a softmax classification head for OT detection or a single linear output for OT height regression. This adaptation allows the model to leverage learned features while being fine-tuned on our specialized dataset.

We trained the models using approximately 10,000 OT cases, which were cropped to a size of 150 × 150 px (km) from HRV scene. All images were scaled to the 0–255 range, where 255 is the maximum reflectance. No correction for sun angle or illumination effects was applied. As is standard practice in neural network training, we divided our dataset into different subsets, with a time-based split.
Training set: Used to update the model's parameters during learning. We allocated 80\% of the data as training set.
Test set: Used to fine-tune hyperparameters and assess model performance during training. This subset comprised 20\% of the data. In order to estimate uncertainty in the model  performance, we used 5-fold cross-validation - training 5 different models for 5 different non-overlapping validation sets, providing multiple estimates of the same score.

\begin{figure}[h]
\centering
\includegraphics[width=0.99\linewidth]{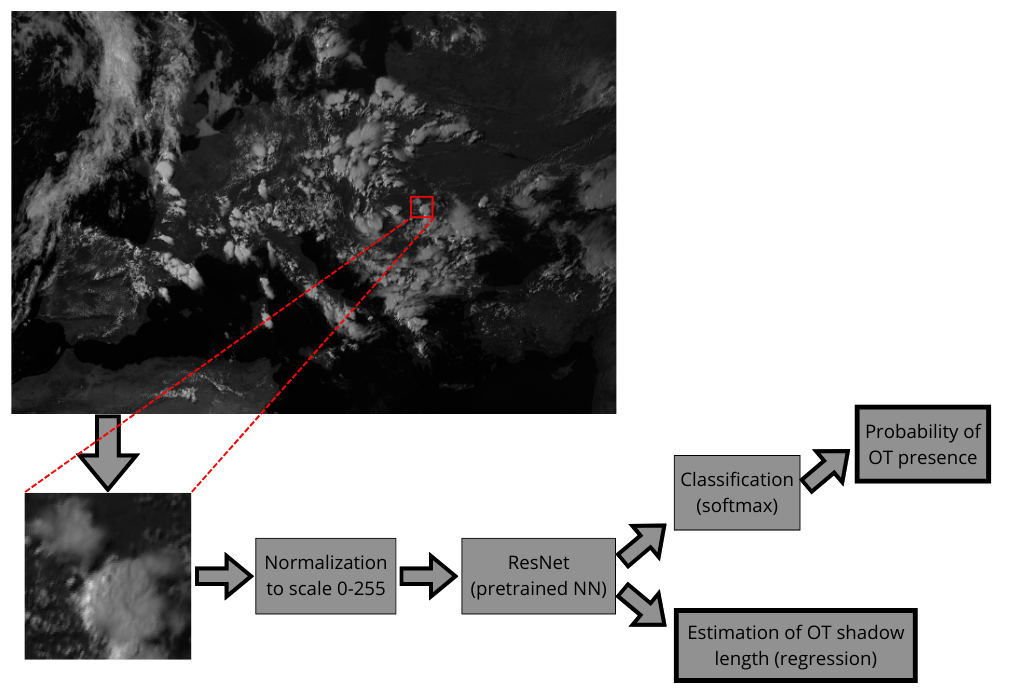}
\caption{The structure of the modeling pipeline includes image cropping from the HRV scene, normalization of pixel values, application of a pretrained ResNet network, and final output through classification or regression, depending on the task.} 
\label{fig:data_flow}
\end{figure}

For the regression task of estimating overshooting top (OT) heights, we used the Mean Squared Error (MSE) as the loss function, which penalizes larger errors more significantly and is well-suited for continuous value prediction. For the classification task of OT detection, we employed the cross-entropy loss, a standard metric for evaluating the performance of classification models. These metrics guided the optimization and evaluation of our neural network models. The complete data processing workflow is illustrated in Figure~\ref{fig:data_flow}.

\subsubsection{Data Augmentation}
One of the main challenges in machine learning is dataset size, as larger datasets generally improve model performance. To enhance the effectiveness of our model, we applied data augmentation \citep{Shorten2019}, a technique that artificially expands the dataset by generating modified versions of existing samples. Augmentation helps mitigate overfitting by improving the model’s ability to generalize beyond the training data. 

The image augmentation techniques \citep{yang2023} include geometric transformations (rotations, flips, scaling and shifts), photometric transformations (adjustments of contrast and brightness) and channel modification (i.e. random channel dropout). These augmentation techniques were carefully selected to preserve the fundamental characteristics of the overshooting tops while ensuring the model was exposed to a diverse range of possible scenarios. By incorporating these transformations, we effectively increased the variability of our training data, helping our model learn more generalizable features from satellite imagery.

We applied the dihedral and rotated augmentations during training. The dihedral transformation randomly applies one of eight possible symmetrical operations (rotations and reflections), while the rotate operation applies a random rotation to the input image. These augmentations help the model generalize better due to variations in image orientation.

By leveraging augmentation, we improved our model's ability to detect and estimate OT heights under various environmental conditions. This step was crucial in compensating for the relatively limited dataset size and ensuring better generalization in real-world applications.

\section{Results}
\label{sec:results}

Our analysis focuses on two tasks: detecting the presence of OTs and estimating their height. While both tasks rely on recognizing patterns in cloud-top images, they differ in methodology. OT detection is a binary classification problem, whereas height estimation is a regression task.

In the following sections, we present the results for each task separately, starting with detection, followed by height estimation. This order reflects their potential future operational use, where detection would precede height estimation.

\subsection{Overshooting Tops Detection}
\label{sec:overshooting_tops_detection}

\begin{figure}[ht!]
\centering
\includegraphics[width=1\linewidth]{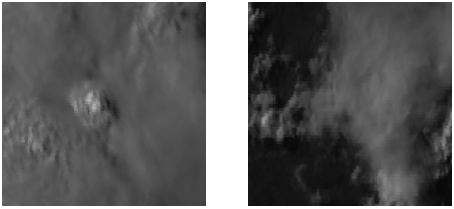}
\caption{Examples of input images used for training. Left: Image centered on an overshooting top (OT). Right: Image without an OT (non-OT case). Both images cover an area of 150 × 150 km.} 
\label{fig:batch}
\end{figure}

The training dataset for OT top detection consists of approximately 10,000 confirmed OT cases and 120,000 non-OT cases, selected from different positions that do not contain any OTs within the same scenes. Images without OT (non-OT type) were cropped to a size of 150 × 150 px (km), in the same way as OT samples. Example of OT and non-OT images from our dataset is in Figure~\ref{fig:batch}. We note, that since many non-OT cases are easy to classify, it is not strictly necessary for the number of OT and non-OT samples in the training dataset to reflect the real-world frequencies exactly. For the training purposes, it is sufficient to include a diverse and representative set of non-OT cases to cover the variety of possible conditions; the exact number of such samples is less critical, as is discussed below.

Many different models were trained with the dataset, with a learning rate of 0.001 for 10 epochs using the method described in section \ref{ml}. Two concepts are important here:
\textbf{Unbalanced Model} – Trained on all data without adjusting for class imbalance, more reflecting real-world conditions where OTs are relatively rare. 
\textbf{Balanced Model} – To address class imbalance, this model applied built-in category balancing during training, ensuring different exposure to both classes. The training parameters remained the same as for the unbalanced model.

\begin{figure}[ht!]
\centering
\includegraphics[width=0.99\linewidth]{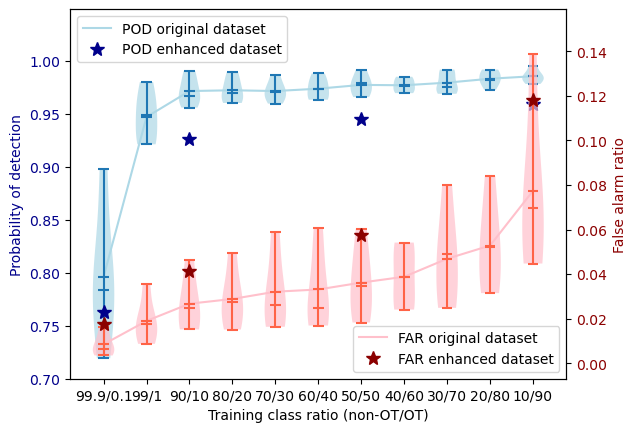}
\caption{Probability of detection (POD, blue, left axis) and false alarm ratio (FAR, red, right axis) as a function of the training class ratio (non-OT/OT). Results are shown for both the original dataset (lines with shading) and the enhanced dataset with additional non-OT samples (stars). Error bars/shaded regions indicate variability across cross-validation runs.} 
\label{fig:bal}
\end{figure}

Data balancing is another reason why it is not necessary to maintain the real-world proportion of non-OT cases during model training. Prior to developing the final model, we trained several models using different ratios of OT and non-OT samples in each training epoch. The performance of these models evaluated using 5-fold cross-validation, particularly in terms of False Alarm Ratio (FAR) and Probability of Detection (POD), is shown in Figure~\ref{fig:bal}. An optimal model should achieve a low FAR - indicating few false alarms - and a high POD - reflecting strong detection capability. Based on these criteria, we selected the balanced model (with a 50/50 ratio of OT to non-OT cases) as the most effective configuration.

\begin{figure}[h]
\centering
\includegraphics[width=0.5\linewidth]{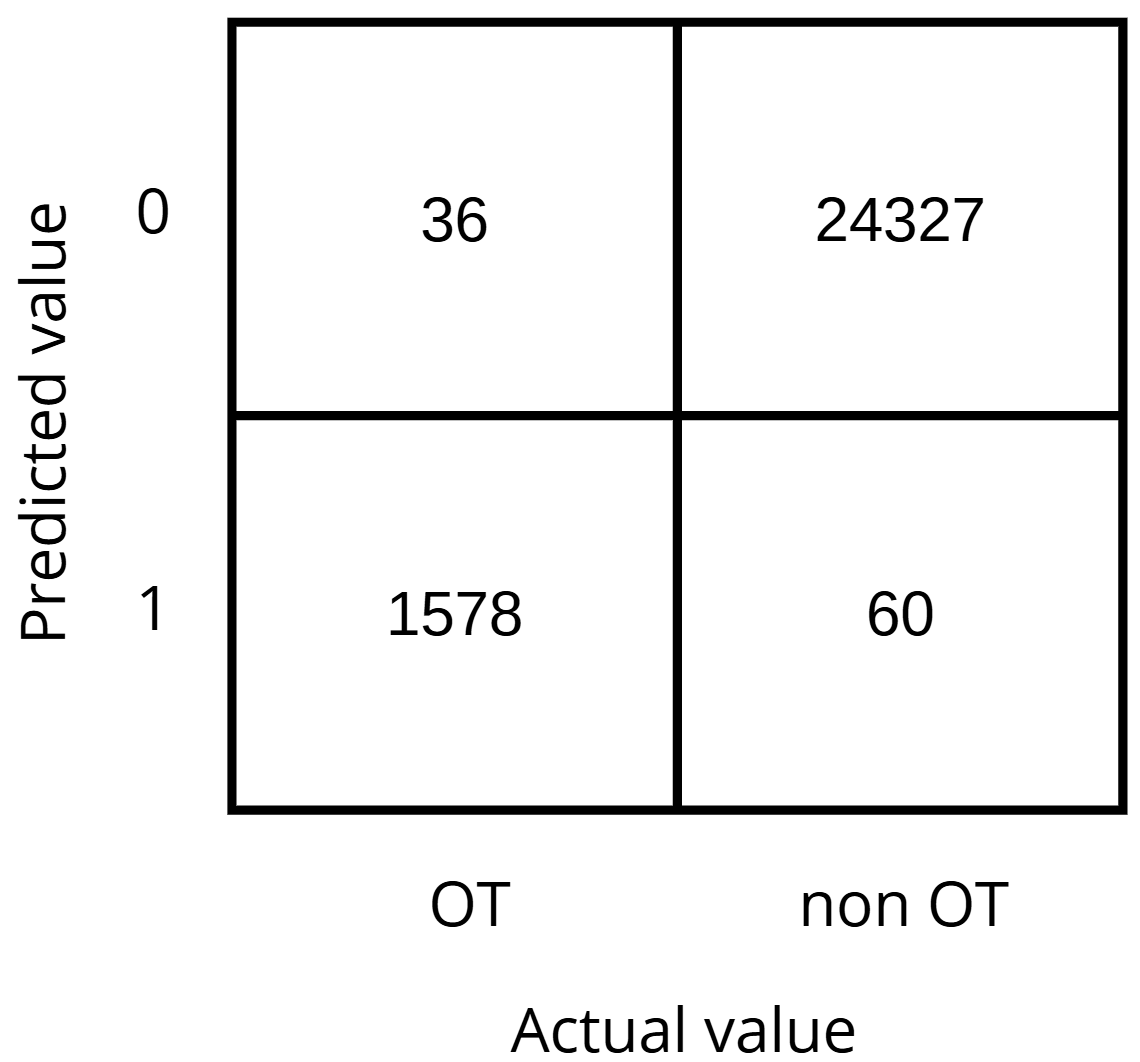}
\caption{Confusion matrix of the 50/50 balanced model evaluated with 5-fold cross-validation (average value).}
\label{fig:cm}
\end{figure}

Performance metrics for the chosen 50/50 balanced model are summarized in Table \ref{score_det}, and its confusion matrix is presented in Fig. \ref{fig:cm}.

The previous experiment reveals the impact of class imbalance during training. In order to further explore the behavior in more realistic settings, where the non-OT/OT ratio will be significantly larger, we generated about  400 000 additional non-OT cases to expand the database, reducing the probability of OT samples in the dataset from $\approx$ 7\% to $\approx$ 1.75\%. Figure \ref{fig:bal} illustrates how the results change when selected models are retrained and tested on the extended database using the same cross-validation procedure as before. As expected, both POD and FAR slightly deteriorated. Nevertheless, these shifts are minor compared to the number of new cases added and should not constraint the model’s applicability in operational practice (see Section \ref{probability_of_OT}).

\begin{table}[h!]
\centering
\begin{tabular}{||c | c||} 
 \hline
 Accuracy & $0.996 \pm 0.001$ \\ 
 \hline
 OT False Alarm Ratio & $0.036 \pm 0.015$ \\
 \hline
 Probability Of OT Detection & $0.977 \pm 0.009$\\
 \hline
\end{tabular}
\caption{Performance scores of the 50/50 balanced model evaluated with 5-fold cross-validation.}
\label{score_det}
\end{table}

\subsubsection{Probability of OT Presence}
\label{probability_of_OT}

\begin{figure}[ht!] 
\centering 
\includegraphics[width=0.99\linewidth]{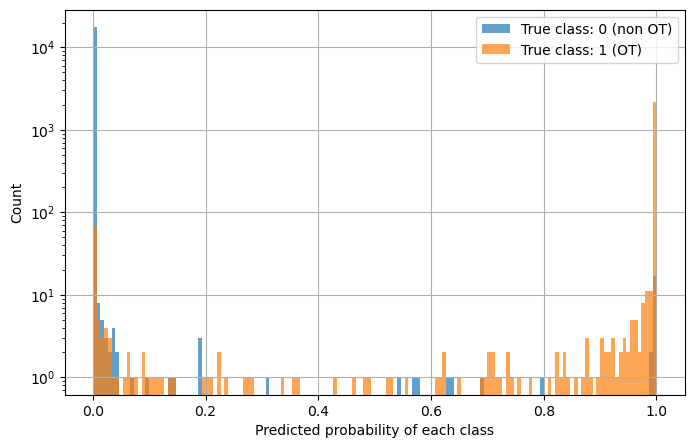} 
\caption{Histogram of predicted probabilities for the balanced classification model. Blue bars represent non-OT cases (true class = 0), and orange bars represent OT cases (true class = 1). The x-axis shows the predicted probability of the image containing an OT, while the y-axis shows the count of predictions on a logarithmic scale.} 
\label{fig:histo} 
\end{figure}

To assess the reliability of the classification model's predicted probabilities, we examined the distribution of output scores on the validation dataset. Figure~\ref{fig:histo} presents a histogram of predicted probabilities for both OT (class 1) and non-OT (class 0) cases. This histogram helps evaluate how well-calibrated the model is—i.e., whether its predicted probabilities reflect actual likelihoods of the respective classes. A well-calibrated model will produce high probabilities for OT cases and low probabilities for non-OT cases, with minimal overlap between the two distributions. In our case, the majority of predictions are confidently close to 0 or 1, indicating that the model makes most of its decisions with high certainty.

\begin{figure}[ht!] 
\centering 
\includegraphics[width=0.99\linewidth]{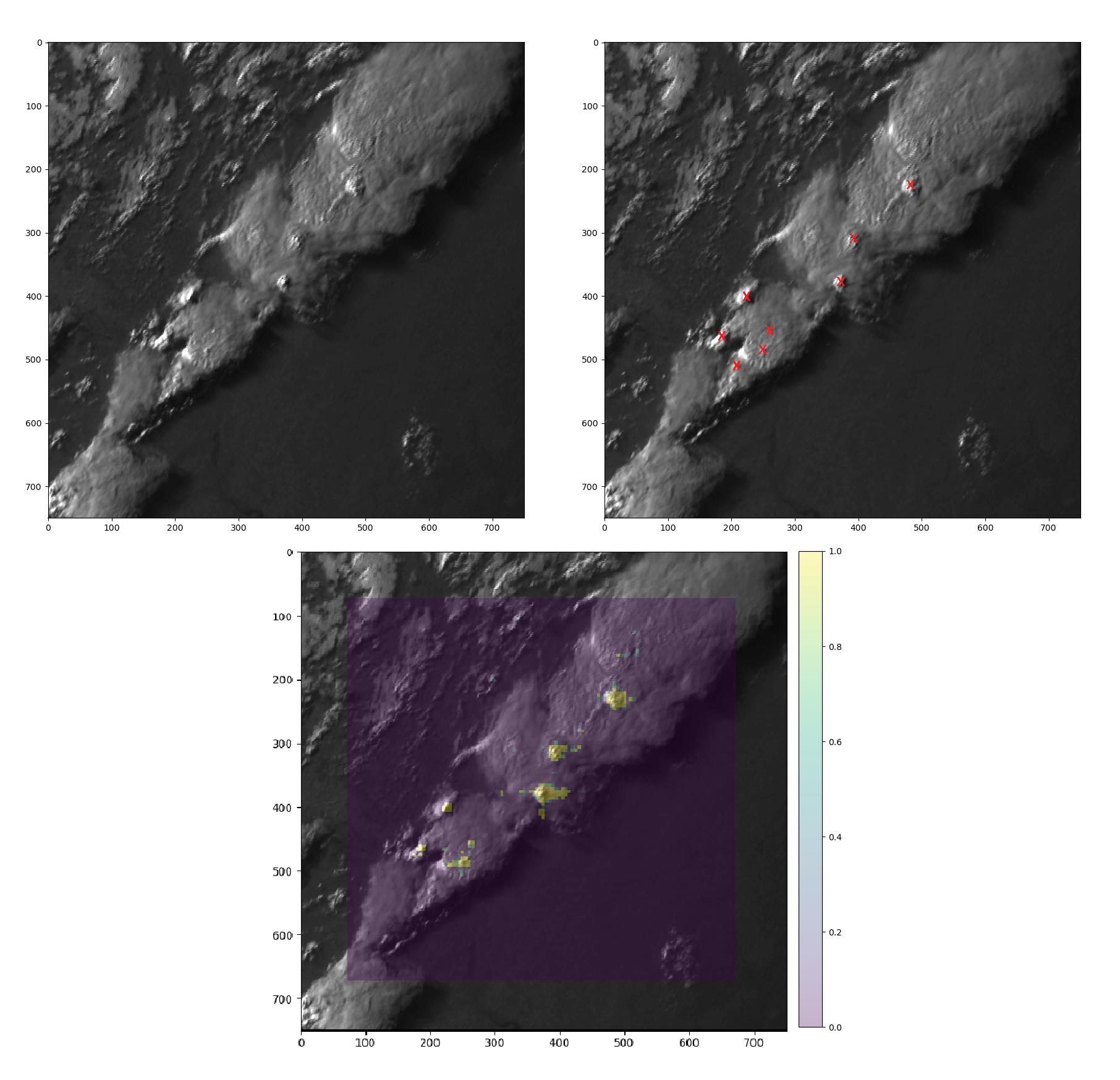} 
\caption{Visualization of OT presence probability across the scene. The image depicts the probability distribution of OT presence (balanced model), with color-coded regions representing different probability levels. The background is an HRV satellite image from August 14, 2014, at 15:45 UTC, centered at 53.3129°N, 32.0929°E. An Azimuthal Equidistant projection was used, with a stride of 5 for product generation. Left: Simple HRV image. Right: HRV image with crosses marking observer-identified OTs—red (ground truth) for certain and orange for uncertain detections. Bottom: Probability of OT presence derived from our model. This case is from the test set that the model did not encounter during training.} 
\label{fig:pro} 
\end{figure}

For operational use, we aim to identify OT locations across the entire scene, rather than just at the central position of the image. To achieve this, we apply the model on many overlapping subparts of the whole scene  simultaneusly, each centered around different pixel. The computed probability is then assigned to the central pixel of the corresponding region. By systematically aggregating these computed values, we construct a complete probability map for the scene. The result of this process is shown in Figure~\ref{fig:pro}.

\subsection{Estimation of the Overshooting Top Height}

A key component of our database is the measurement of OT height. This information allows us to develop a model for estimating this parameter from the imagery. In this case, the task is a regression problem, as we predict continuous values rather than categorical labels.

\begin{figure}[h]
\centering
\includegraphics[width=0.99\linewidth]{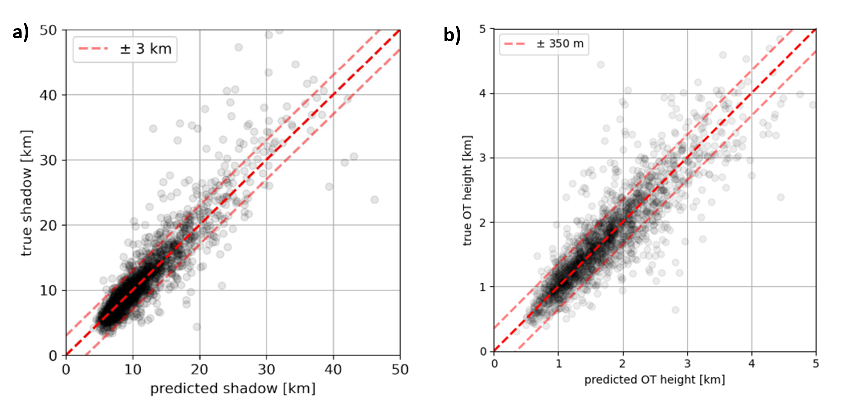} 

\caption{Comparison between predicted and actual values of OT shadow lengths (a) and OT heights (b).}
\label{fig:image2}
\end{figure}

The model was trained utilizing a ResNet-18 architecture fine-tuned with a learning rate of 0.01. The training dataset was prepared as described in Section \ref{ml}. The different error metrics for validation set of model performance are presented in Table \ref{table2}. Figure \ref{fig:image2}a compares the predicted shadow lengths with the true values from the database, while Figure \ref{fig:image2}b shows a similar comparison for OT heights. The model outputs shadow lengths, and the OT heights are subsequently computed based on the geometric approach illustrated in Figure~\ref{fig:geometry}.

\begin{table}[h!]
\centering
\begin{tabular}{||c | c | c||} 
 \hline
 Type of Score & Shadow Length [km] & OT Height [m] \\ [0.5ex] 
 \hline\hline
 Median Error & 1.1 & 174 \\ 
 \hline
 Mean Error & 1.8 & 251 \\
 \hline
 RMSE & 3.1 & 369 \\ 
 \hline
\end{tabular}
\caption{Performance metrics for shadow length and OT height predictions.}
\label{table2}
\end{table}

\subsubsection{OT Shadow Mask}

\begin{figure}[ht!]
\centering
\includegraphics[width=0.99\linewidth]{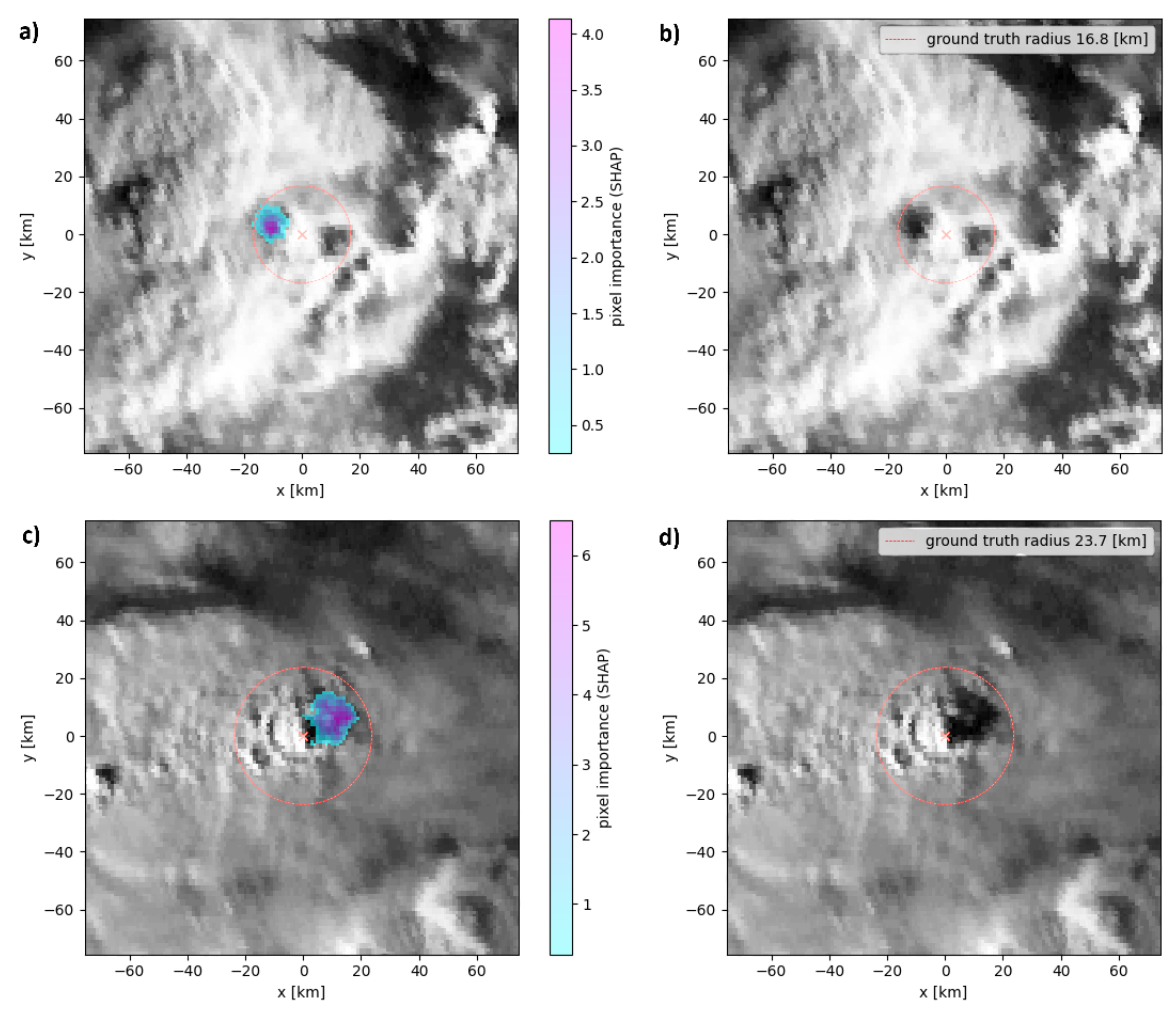}
\caption{Visualization of OT shadow prediction using SHAP values and corresponding HRV imagery. Panels (a) and (c) show pixel importance maps based on SHAP values, highlighting areas that significantly influence the predicted OT shadow length. Panels (b) and (d) display the same HRV images as (a) and (c), respectively, with the OT shadow indicated. The red circles in all panels represent radii equal to the ground truth shadow lengths.}
\label{fig:shapfigure}
\end{figure}

Our novel approach for detecting OT shadows is based on image recognition using HRV images and convolutional neural networks (CNNs). With the model that estimates OT shadow length, we can also indirectly identify pixel mask of the OT shadow, by assessing which pixel(s) contribute most significantly to the length estimation. This process falls under the domain of explainable machine learning. As the length of the shadow is approximately proportional to the number of pixels in the shadow, we propose to employ a SHAP (Shapley Additive Explanations) \citep{shap} statistical technique that decomposes model prediction (shadow length) into the sum of contributions from individual input features (pixel values). SHAP, therefore, assigns an importance score to each pixel, indicating its contribution to the predicted shadow length. Pixels with a high impact (where the shadow is present) will have nonzero SHAP values, while those with little to no influence will have SHAP values close to zero or negative. By thresholding the SHAP values one can consequently obtain a pixel mask of the shadow as illustrated in Figure \ref{fig:shapfigure}. The largest continuous patch can then be interpreted as the OT shadow mask.

\subsection{Comparison with Other Methods}

Automatic detection of OTs from satellite data can be categorized into traditional methods, which do not use machine learning (ML), and modern ML-based approaches. Another distinguishing factor is the type of satellite channel used. Most methods rely on infrared (IR) channels, particularly brightness temperature, while visible channels can also be utilized, as they reveal cloud-top structures.

Traditional IR-based methods include those described in \citet{bedka2010}. A basic approach involves setting a brightness temperature threshold of 215K, where temperatures below this value indicate potential OTs. More advanced methods include "IRW texture" and "WV-IRW BTD," which leverage differences between the water vapor channel ($\approx 7\mu m$) and the IR channel in the atmospheric window ($\approx 11\mu m$) \citep{bedka2010}. However, these methods struggle in cases known as "BTD anomalies," where the lowest brightness temperature in the IR channel does not align with the highest WV-IRW difference \citep{setv2007}. The validation of these approaches has been conducted in studies such as \citet{bedka2012} and \citet{mikus}. Additionally, \citet{rad2015} tested these methods on their dataset, with the best-performing method (O3-IRW) detecting only 61\% of OTs.

More recently, ML-based methods have been explored. The most common approach involves convolutional neural networks (CNNs), typically incorporating both visible and IR satellite data. For instance, \citet{kim2018} used data from the Himawari satellite, which primarily covers the East Asia tropics, training their model on 10,000 samples containing 571 OT cases and achieving a probability of detection (POD) of 79.68\%. Similarly, \citet{kan2020} utilized GOES satellite data, albeit with a smaller dataset of 459 images, and obtained a POD of 79.31\%. Both studies used 31×31 pixel patches for OT detection, with the OT positioned at the center of the image.

Compared to these previous works, our approach has several key differences. First, our dataset is significantly larger, containing approximately 120,000 non-OT cases and around 10,000 OT cases. A larger dataset enhances statistical reliability and model robustness. Additionally, our model achieves a POD reaching 97.7\%, surpassing previous studies. However, it is important to note that our dataset is focused exclusively on Europe, a relatively small and homogeneous region. This regional specificity likely contributes to improved performance within Europe, but the model's accuracy outside this area may be lower.

Another key difference lies in the exclusive use of the visible channel. Despite this limitation, we were able to develop a reliable model; however, it is restricted to daytime observations of OTs with a shadow (see Figure~\ref{fig:sunangle}). In future work, we plan to incorporate infrared (IR) channels to enable detection during nighttime and low-light conditions.

An important aspect of our work is the creation of a prototype product based on this model. This demonstrates the model's potential for operational use and highlights its practical applicability in real-time meteorological monitoring.

\section{Conclusion}

We developed machine learning models for the automatic detection of overshooting tops (OTs) and the estimation of their heights from visible satellite imagery. Both models are based on convolutional neural networks (CNNs), utilizing a pretrained ResNet architecture and trained on a large database of manually detected and measured OTs.

For the task of OT detection, multiple models were trained using varying non-OT/OT ratios in the training dataset. The results demonstrate that an increasing proportion of OT cases leads to a systematic rise in both the probability of detection (POD) and the false alarm ratio (FAR). This relationship suggests that models trained with a high prevalence of non-OT cases adopt a more conservative detection strategy, yielding lower FAR values but at the expense of reduced POD. Such models identify OTs only under conditions of high confidence, which results in a substantial number of missed detections. Conversely, models trained with a higher representation of OT cases exhibit greater sensitivity but at the cost of an increased number of false alarms. Considering these trade-offs, the balanced 50/50 ratio was identified as the most effective compromise, achieving a favorable balance between POD 97.7\% and FAR 3.6\%. This configuration enhances the model’s practical applicability, as it provides reliable detection performance while maintaining operational relevance by highlighting regions of potential OT activity for subsequent expert evaluation. 

The presented scores reflect the model’s performance on the validation set, which is based on the implicit definition of what constitutes an OT, determined during the assembly of the training database in \citep{kanak2012} by its authors, and on the procedure for generating the non-OT portion of the database described in Section \ref{sec:overshooting_tops_detection}. The probability of detection quantifies the capacity of the model to identify similar OTs as appear in the database but in new situations, while the false alarm rate estimates the lower bound on the occurrence of incorrect OT identifications. In operational use, the metrics may differ due to (i) the subjectivity of the OT definition (expert opinions often diverge, especially for small OTs), (ii) the occurrence of out-of-training-distribution cases, and (iii) a different ratio of OT to non-OT cases in real-world situations. The first two issues could be mitigated by further expanding the training database and by employing soft labels that capture the degree of certainty of multiple experts for each sample. To address the third issue, we tested the performance of the model for varying ratios of OT and non-OT cases in both the training and validation sets, as shown in Figure~\ref{fig:bal}. The results indicate a degradation in scores with an increasing number of non-OT validation samples. However, this decrease is not proportional to the change in ratio of the classes, as most non-OT samples are, in general, relatively easy to classify. Consequently, the performance of the model is expected to remain reasonably high even under operational conditions where non-OT cases dominate, limited mainly by the range of suitable Sun zenith angles.

The OT height estimation model achieved a mean error of 1.8 km in shadow length measurements, which translates to 250 m in OT height estimation. Given that the resolution of HRV data is approximately 1 km at nadir, this error is comparable to the uncertainty inherent in manual measurements, which can vary depending on the observer's expertise. A relatively large error was observed for the OTs with the largest shadow lengths. This is likely due to their infrequent occurrence in the dataset and is expected to improve with dataset size.

Compared to previous studies, our approach introduces two key advancements. First, we rely exclusively on the high-resolution visible (HRV) channel, making the model particularly useful for identifying anomalous cases that do not follow standard brightness temperature patterns. Second, we not only detect OTs but also estimate their height, which serves as an important indicator of strong updrafts and can contribute to future assessments of storm severity and possibly predictive models based on OT occurrence. Future work could explore integrating OT detection and height estimation into a unified framework and incorporating thermal data.   

Our model is also adaptable for use with data from polar-orbiting satellites and the Meteosat Third Generation (MTG), as long as the visible channel imagery at least roughly meets the 1 km resolution requirement. Since many polar satellites offer even higher resolutions and MTG data can be appropriately resampled, this adaptation is entirely feasible. The model was trained on data projected to a uniform 1 km resolution, which facilitates compatibility with other satellite sources. This approach supports transfer learning, allowing the model to be fine-tuned for different satellites or extended by incorporating a variable that specifies the satellite type.

In summary, we have developed a robust and versatile machine learning approach for detecting overshooting tops and estimating their heights using only visible satellite imagery. Despite the limitations of visible data availability during nighttime and lack of shadows for large Sun zenith angles, the model demonstrates high accuracy and operational potential, with adaptability to different satellite systems through transfer learning. This work provides a foundation for developing more advanced nowcasting tools, and future integration with infrared data is expected to further improve its reliability.

\section*{CRediT authorship contribution statement}
Anežka Doležalová: Writing – original draft, Writing – review \& editing, Formal analysis, Software, Visualization. Jakub Seidl: Writing – review \& editing, Supervision, Software, Visualization, Conceptualization. Jindřich Šťástka: Writing – review \& editing, Supervision, Methodology, Formal analysis, Validation. Ján Kaňák: Supervision, Methodology, Conceptualization, Validation.

\section*{Acknowledgements}

This work was supported by the Department of Atmospheric Physics, Faculty of Mathematics and Physics, Charles University, particularly through the Charles University project SVV 260825 and access to its computational facilities.
Part of this study was carried out with the support of the Czech Ministry of Environment, DKRVO, CHMI 2023-2027 program.
Part of the computational resources were provided by the e-INFRA CZ project (ID:90254), supported by the Ministry of Education, Youth and Sports of the Czech Republic.

\section*{Declaration of generative AI}

During the preparation of this work the authors used ChatGPT in order to improve language and readability. After using this tool/service, the authors reviewed and edited the content as needed and take full responsibility for the content of the publication.

\bibliographystyle{elsarticle-harv} 
\bibliography{sample}

\begin{thebibliography}{24}
\expandafter\ifx\csname natexlab\endcsname\relax\def\natexlab#1{#1}\fi
\providecommand{\url}[1]{\texttt{#1}}
\providecommand{\href}[2]{#2}
\providecommand{\path}[1]{#1}
\providecommand{\DOIprefix}{doi:}
\providecommand{\ArXivprefix}{arXiv:}
\providecommand{\URLprefix}{URL: }
\providecommand{\Pubmedprefix}{pmid:}
\providecommand{\doi}[1]{\href{http://dx.doi.org/#1}{\path{#1}}}
\providecommand{\Pubmed}[1]{\href{pmid:#1}{\path{#1}}}
\providecommand{\bibinfo}[2]{#2}
\ifx\xfnm\relax \def\xfnm[#1]{\unskip,\space#1}\fi
\bibitem[{Bedka et~al.(2010)Bedka, Brunner, Dworak, Feltz, Otkin and T.}]{bedka2010}
\bibinfo{author}{Bedka, K.}, \bibinfo{author}{Brunner, J.}, \bibinfo{author}{Dworak, R.}, \bibinfo{author}{Feltz, W.}, \bibinfo{author}{Otkin, J.}, \bibinfo{author}{T., G.}, \bibinfo{year}{2010}.
\newblock \bibinfo{title}{Objective satellite-based detection of overshooting tops using infrared window channel brightness temperature gradients}.
\newblock \bibinfo{journal}{Journal of Applied Meteorology and Climatology} , \bibinfo{pages}{181--202}\URLprefix \url{https://doi.org/10.1175/2009JAMC2286.1}.
\bibitem[{Bedka et~al.(2012)Bedka, Dworak, Brunner and Feltz}]{bedka2012}
\bibinfo{author}{Bedka, K.M.}, \bibinfo{author}{Dworak, R.}, \bibinfo{author}{Brunner, J.}, \bibinfo{author}{Feltz, W.}, \bibinfo{year}{2012}.
\newblock \bibinfo{title}{Validation of satellite-based objective overshooting cloud-top detection methods using cloudsat cloud profiling radar observations.}
\newblock \bibinfo{journal}{Journal of Applied Meteorology and Climatology} , \bibinfo{pages}{1811--1822}\URLprefix \url{https://doi.org/10.1175/JAMC-D-11-0131.1}.
\bibitem[{Bluestein et~al.(2019)Bluestein, Lindsey, Bikos, Reif and Wienhoff}]{Bluestein2019}
\bibinfo{author}{Bluestein, H.B.}, \bibinfo{author}{Lindsey, D.T.}, \bibinfo{author}{Bikos, D.}, \bibinfo{author}{Reif, D.W.}, \bibinfo{author}{Wienhoff, Z.B.}, \bibinfo{year}{2019}.
\newblock \bibinfo{title}{The relationship between overshooting tops in a tornadic supercell and its radar-observed evolution}.
\newblock \bibinfo{journal}{Monthly Weather Review} \bibinfo{volume}{147}, \bibinfo{pages}{4151 -- 4176}.
\newblock \URLprefix \url{https://journals.ametsoc.org/view/journals/mwre/147/11/mwr-d-19-0159.1.xml}, \DOIprefix\doi{10.1175/MWR-D-19-0159.1}.
\bibitem[{Deng et~al.(2009)Deng, Dong, Socher, Li, Li and Fei-Fei}]{imagenet}
\bibinfo{author}{Deng, J.}, \bibinfo{author}{Dong, W.}, \bibinfo{author}{Socher, R.}, \bibinfo{author}{Li, L.J.}, \bibinfo{author}{Li, K.}, \bibinfo{author}{Fei-Fei, L.}, \bibinfo{year}{2009}.
\newblock \bibinfo{title}{Imagenet: A large-scale hierarchical image database}, in: \bibinfo{booktitle}{2009 IEEE Conference on Computer Vision and Pattern Recognition}, pp. \bibinfo{pages}{248--255}.
\newblock \DOIprefix\doi{10.1109/CVPR.2009.5206848}.
\bibitem[{Dworak et~al.(2012)Dworak, Bedka, Brunner and Feltz}]{dworak2012}
\bibinfo{author}{Dworak, R.}, \bibinfo{author}{Bedka, K.}, \bibinfo{author}{Brunner, J.}, \bibinfo{author}{Feltz, W.}, \bibinfo{year}{2012}.
\newblock \bibinfo{title}{{Comparison between GOES-12 Overshooting-Top Detections, WSR-88D Radar Reflectivity, and Severe Storm Reports}}.
\newblock \bibinfo{journal}{Weather and Forecasting} \bibinfo{volume}{27}.
\newblock \DOIprefix\doi{10.1175/WAF-D-11-00070.1}.
\bibitem[{He et~al.(2016)He, Zhang, Ren and Sun}]{resnet}
\bibinfo{author}{He, K.}, \bibinfo{author}{Zhang, X.}, \bibinfo{author}{Ren, S.}, \bibinfo{author}{Sun, J.}, \bibinfo{year}{2016}.
\newblock \bibinfo{title}{Deep residual learning for image recognition}.
\newblock \bibinfo{journal}{In Proceedings of the IEEE conference on computer vision and pattern recognition} , \bibinfo{pages}{770--778}.
\bibitem[{Howard and Gugger(2021)}]{Howard_Gugger_2021}
\bibinfo{author}{Howard, J.}, \bibinfo{author}{Gugger, S.}, \bibinfo{year}{2021}.
\newblock \bibinfo{title}{Deep learning for coders with FASTAI and pytorch: AI applications without a Phd}.
\newblock \bibinfo{publisher}{O’Reilly Media}.
\bibitem[{Kanneganti(2020)}]{kan2020}
\bibinfo{author}{Kanneganti, G.T.}, \bibinfo{year}{2020}.
\newblock \bibinfo{title}{Detection of Overshooting Cloud Tops with Convolutional Neural Networks}.
\newblock \bibinfo{type}{Master's thesis}. University of Oklahoma.
\newblock \bibinfo{note}{Available at \url{https://hdl.handle.net/11244/324409}}.
\bibitem[{Kaňák et~al.(2012)Kaňák, Bedka and A.}]{kanak2012}
\bibinfo{author}{Kaňák, J.}, \bibinfo{author}{Bedka, K.M.}, \bibinfo{author}{A., S.}, \bibinfo{year}{2012}.
\newblock \bibinfo{title}{Mature convective storms and their overshooting tops over central europe: Overshooting top height analysis for summers 2009-2011}.
\newblock \bibinfo{journal}{Conference: 2012 EUMETSAT Meteorological Satellite Conference, Session 7} \URLprefix \url{https://www.eumetsat.int/media/8672}.
\bibitem[{Kim et~al.(2018)Kim, Lee and Im}]{kim2018}
\bibinfo{author}{Kim, M.}, \bibinfo{author}{Lee, J.}, \bibinfo{author}{Im, J.}, \bibinfo{year}{2018}.
\newblock \bibinfo{title}{Deep learning-based monitoring of overshooting cloud tops from geostationary satellite data}.
\newblock \bibinfo{journal}{GIScience \& Remote Sensing} \bibinfo{volume}{55}, \bibinfo{pages}{763--792}.
\newblock \URLprefix \url{https://doi.org/10.1080/15481603.2018.1457201}, \DOIprefix\doi{10.1080/15481603.2018.1457201}.
\bibitem[{Kingma and Ba(2017)}]{kingma2017}
\bibinfo{author}{Kingma, D.P.}, \bibinfo{author}{Ba, J.}, \bibinfo{year}{2017}.
\newblock \bibinfo{title}{Adam: A method for stochastic optimization}.
\newblock \URLprefix \url{https://arxiv.org/abs/1412.6980}, \href{http://arxiv.org/abs/1412.6980}{{\tt arXiv:1412.6980}}.
\bibitem[{Lundberg and Lee(2017)}]{shap}
\bibinfo{author}{Lundberg, S.M.}, \bibinfo{author}{Lee, S.I.}, \bibinfo{year}{2017}.
\newblock \bibinfo{title}{A unified approach to interpreting model predictions}, in: \bibinfo{booktitle}{Advances in Neural Information Processing Systems}, \bibinfo{publisher}{Curran Associates, Inc.}
\bibitem[{Mikuš and Strelec~Mahović(2013)}]{mikus}
\bibinfo{author}{Mikuš, P.}, \bibinfo{author}{Strelec~Mahović, N.}, \bibinfo{year}{2013}.
\newblock \bibinfo{title}{Satellite-based overshooting top detection methods and an analysis of correlated weather conditions}.
\newblock \bibinfo{journal}{Atmospheric Research} \bibinfo{volume}{123}, \bibinfo{pages}{268--280}.
\newblock \URLprefix \url{https://www.sciencedirect.com/science/article/pii/S0169809512002980}, \DOIprefix\doi{https://doi.org/10.1016/j.atmosres.2012.09.001}. \bibinfo{note}{6th European Conference on Severe Storms 2011. Palma de Mallorca, Spain}.
\bibitem[{Piasecki et~al.(2023)Piasecki, Matczak, Taszarek, Czernecki, Skop and Sobisiak}]{pia}
\bibinfo{author}{Piasecki, K.}, \bibinfo{author}{Matczak, P.}, \bibinfo{author}{Taszarek, M.}, \bibinfo{author}{Czernecki, B.}, \bibinfo{author}{Skop, F.}, \bibinfo{author}{Sobisiak, A.}, \bibinfo{year}{2023}.
\newblock \bibinfo{title}{Giant hail in poland produced by a supercell merger in extreme instability – a sign of a warming climate?}
\newblock \bibinfo{journal}{Atmospheric Research} \bibinfo{volume}{292}, \bibinfo{pages}{106843}.
\newblock \URLprefix \url{https://www.sciencedirect.com/science/article/pii/S0169809523002405}, \DOIprefix\doi{https://doi.org/10.1016/j.atmosres.2023.106843}.
\bibitem[{Radová(2015)}]{rad2015}
\bibinfo{author}{Radová, M.}, \bibinfo{year}{2015}.
\newblock \bibinfo{title}{Cloud-top morphology of convective storms as observed by meteorological satellites}.
\newblock \bibinfo{note}{Available at \url{http://hdl.handle.net/20.500.11956/76186}}.
\bibitem[{Ricchi et~al.(2023)Ricchi, Rotunno, Miglietta, Picciotti, Montopoli, Marzano, Baldini, Vulpiani, Tiesi and Ferretti}]{RICCHI}
\bibinfo{author}{Ricchi, A.}, \bibinfo{author}{Rotunno, R.}, \bibinfo{author}{Miglietta, M.M.}, \bibinfo{author}{Picciotti, E.}, \bibinfo{author}{Montopoli, M.}, \bibinfo{author}{Marzano, F.}, \bibinfo{author}{Baldini, L.}, \bibinfo{author}{Vulpiani, G.}, \bibinfo{author}{Tiesi, A.}, \bibinfo{author}{Ferretti, R.}, \bibinfo{year}{2023}.
\newblock \bibinfo{title}{Analysis of the development mechanisms of a large-hail storm event on the adriatic sea}.
\newblock \bibinfo{journal}{Atmospheric Research} \bibinfo{volume}{296}, \bibinfo{pages}{107079}.
\newblock \URLprefix \url{https://www.sciencedirect.com/science/article/pii/S0169809523004763}, \DOIprefix\doi{https://doi.org/10.1016/j.atmosres.2023.107079}.
\bibitem[{Schmetz et~al.(2002)Schmetz, Pili, Tjemkes, Just, Kerkmann, Rota and Ratier}]{schmetz2002}
\bibinfo{author}{Schmetz, J.}, \bibinfo{author}{Pili, P.}, \bibinfo{author}{Tjemkes, S.}, \bibinfo{author}{Just, D.}, \bibinfo{author}{Kerkmann, J.}, \bibinfo{author}{Rota, S.}, \bibinfo{author}{Ratier, A.}, \bibinfo{year}{2002}.
\newblock \bibinfo{title}{An introduction to meteosat second generation (msg)}.
\newblock \bibinfo{journal}{Bull. Amer. Meteor. Soc.} \bibinfo{volume}{83}, \bibinfo{pages}{977--992}.
\newblock \DOIprefix\doi{\texttt{https://doi.org/10.1175/1520-0477(2002)083<0977:AITMSG>2.3.CO;2}}.
\bibitem[{Setvák et~al.(2010)Setvák, Lindsey, Novák and Wang}]{setv2010}
\bibinfo{author}{Setvák, M.}, \bibinfo{author}{Lindsey, D.T.}, \bibinfo{author}{Novák, P.}, \bibinfo{author}{Wang, P.K.}, \bibinfo{year}{2010}.
\newblock \bibinfo{title}{Satellite-observed cold-ring-shaped features atop deep convective clouds}.
\newblock \bibinfo{journal}{Atmospheric Research} , \bibinfo{pages}{80--96}\URLprefix \url{https://doi.org/10.1016/j.atmosres.2010.03.009}.
\bibitem[{Setvák et~al.(2007)Setvák, Rabin and Wang}]{setv2007}
\bibinfo{author}{Setvák, M.}, \bibinfo{author}{Rabin, R.M.}, \bibinfo{author}{Wang, P.K.}, \bibinfo{year}{2007}.
\newblock \bibinfo{title}{Contribution of the modis instrument to observations of deep convective storms and stratospheric moisture detection in goes and msg imagery.}
\newblock \bibinfo{journal}{Atmospheric Research} , \bibinfo{pages}{505--518}\URLprefix \url{10.1016/j.atmosres.2005.09.015}.
\bibitem[{Shorten and Khoshgoftaar(2019)}]{Shorten2019}
\bibinfo{author}{Shorten, C.}, \bibinfo{author}{Khoshgoftaar, T.M.}, \bibinfo{year}{2019}.
\newblock \bibinfo{title}{A survey on image data augmentation for deep learning}.
\newblock \bibinfo{journal}{Journal of Big Data} \bibinfo{volume}{6}, \bibinfo{pages}{2196--1115}.
\newblock \URLprefix \url{https://doi.org/10.1186/s40537-019-0197-0}, \DOIprefix\doi{10.1186/s40537-019-0197-0}.
\bibitem[{Sun et~al.(2024)Sun, Lu and Fu}]{sun2024}
\bibinfo{author}{Sun, N.}, \bibinfo{author}{Lu, G.}, \bibinfo{author}{Fu, Y.}, \bibinfo{year}{2024}.
\newblock \bibinfo{title}{Microphysical characteristics of precipitation within convective overshooting over east china observed by {GPM DPR and ERA5}}.
\newblock \bibinfo{journal}{Atmospheric Chemistry and Physics} \bibinfo{volume}{24}, \bibinfo{pages}{7123--7135}.
\newblock \URLprefix \url{https://acp.copernicus.org/articles/24/7123/2024/}, \DOIprefix\doi{10.5194/acp-24-7123-2024}.
\bibitem[{Weiss and Khoshgoftaar(2016)}]{weiss2016}
\bibinfo{author}{Weiss, K.}, \bibinfo{author}{Khoshgoftaar, Taghi M.~andWang, D.}, \bibinfo{year}{2016}.
\newblock \bibinfo{title}{A survey of transfer learning}.
\newblock \bibinfo{journal}{Journal of Big Data} \bibinfo{volume}{3}, \bibinfo{pages}{2196--1115}.
\newblock \URLprefix \url{https://doi.org/10.1186/s40537-016-0043-6}, \DOIprefix\doi{10.1186/s40537-016-0043-6}.
\bibitem[{Yair et~al.(2024)Yair, Korzets, Devir, Korman and Stibbe}]{YAIR}
\bibinfo{author}{Yair, Y.}, \bibinfo{author}{Korzets, M.}, \bibinfo{author}{Devir, A.}, \bibinfo{author}{Korman, M.}, \bibinfo{author}{Stibbe, E.}, \bibinfo{year}{2024}.
\newblock \bibinfo{title}{Space-based optical imaging of blue corona discharges on a cumulonimbus cloud top}.
\newblock \bibinfo{journal}{Atmospheric Research} \bibinfo{volume}{305}, \bibinfo{pages}{107445}.
\newblock \URLprefix \url{https://www.sciencedirect.com/science/article/pii/S0169809524002278}, \DOIprefix\doi{https://doi.org/10.1016/j.atmosres.2024.107445}.
\bibitem[{Yang et~al.(2023)Yang, Xiao, Zhang, Guo, Zhao and Shen}]{yang2023}
\bibinfo{author}{Yang, S.}, \bibinfo{author}{Xiao, W.}, \bibinfo{author}{Zhang, M.}, \bibinfo{author}{Guo, S.}, \bibinfo{author}{Zhao, J.}, \bibinfo{author}{Shen, F.}, \bibinfo{year}{2023}.
\newblock \bibinfo{title}{Image data augmentation for deep learning: A survey}.
\newblock \URLprefix \url{https://arxiv.org/abs/2204.08610}, \href{http://arxiv.org/abs/2204.08610}{{\tt arXiv:2204.08610}}.

\end{thebibliography}

\end{document}